\documentstyle[11pt,newpasp,psfig,epsf,twoside]{article}
\def\gtsim{{_>\atop{^\sim}}}
\def\ltsim{{_<\atop{^\sim}}}

\begin{document}

\title{Chemistry in the Envelopes around Massive Young Stars}
\author{Ewine F.\ van Dishoeck and Floris F.S.\ van der Tak}
\affil{Leiden Observatory, P.O.\ Box 9513, 2300 RA Leiden,
The Netherlands}

\begin{abstract}
Recent observational studies of intermediate- and high-mass
 star-forming regions at submillimeter and infrared wavelengths are
 reviewed, and chemical diagnostics of the different physical
 components associated with young stellar objects are summarized.
 Procedures for determining the temperature, density and abundance
 profiles in the envelopes are outlined. A detailed study of a set of
 infrared-bright massive young stars reveals systematic increases in
 the gas/solid ratios, the abundances of evaporated molecules, and the
 fraction of heated ices with increasing temperature. Since these
 diverse phenomena involve a range of temperatures from $<100\,$K to
 1000~K, the enhanced temperatures must be communicated to both the
 inner and outer parts of the envelopes. This `global heating'
 plausibly results from the gradual dispersion of the envelopes with
 time. Similarities and differences with low-mass YSOs are discussed.
 The availability of accurate physical models will allow chemical
 models of ice evaporation followed by `hot core' chemistry to be
 tested in detail.

\end{abstract}


\section{Introduction}

Massive star-forming regions have traditionally been prime targets for
astrochemistry owing to their bright molecular lines (e.g., Johansson
et al.\ 1984, Cummins et al.\ 1986, Irvine et al.\ 1987, Ohishi 1997).
Massive young stellar objects (YSOs) have luminosities of $\sim 10^4 -
10^6$ L$_\odot$ and involve young O- and B-type stars.  Because their
formation proceeds more rapidly than that of low-mass stars and
involves ionizing radiation, substantial chemical differences may be
expected. The formation of high mass stars is much less well
understood than that of low-mass stars. For
example, observational phenomena such as ultracompact H II regions,
hot cores, masers and outflows have not yet been linked into a single
evolutionary picture.  Chemistry may well be an important diagnostic
tool in establishing such a sequence.
 
Most of the early work on massive star-forming regions has centered on
two sources, Orion--KL and SgrB2. Numerous line surveys at millimeter
(e.g., Blake et al.\ 1987, Turner 1991) and submillimeter (Jewell et
al.\ 1989, Sutton et al.\ 1991, 1995, Schilke et al.\ 1997)
wavelengths have led to an extensive inventory of molecules through
identification of thousands of lines. In addition, the surveys have
shown strong chemical variations between different sources.

In recent years, new observational tools have allowed a more detailed and
systematic study of the envelopes of massive YSOs.  Submillimeter observations
routinely sample smaller beams (typically 15$''$ vs.\ 30$''$--1$'$ ) and
higher critical densities ($\geq 10^6$ vs.\ $10^4$ cm$^{-3}$) than the earlier
work.  Moreover, interferometers at 3 and 1 millimeter provide maps with
resolutions of $0.5''$--5$''$. Finally, ground- and space-based infrared
observations allow both the gas and the ices to be sampled (e.g., Evans et
al.\ 1991, van Dishoeck et al.\ 1999). These observational developments have
led to a revival of the study of massive star formation within the last few
years.  Recent overviews of the physical aspects of high-mass star formation
are found in Churchwell (1999) and Garay \& Lizano (1999).

In this brief review, we will first summarize available observational
diagnostics to study the different phases and physical components
associated with massive star formation.  Subsequently, an overview of
recent results on intermediate mass YSOs is given, which are often
better characterized than their high-mass counterparts because of
their closer distance. Subsequently, we will discuss a specific sample
of embedded massive YSOs which have been studied through a combination
of infrared and submillimeter data. After illustrating
the modeling techniques, we address the question how the
observed chemical variations are related to evolutionary effects,
different conditions in the envelope (e.g., $T$, mass) or different
luminosities of the YSOs.  More extensive overviews of the chemical
evolution of star-forming regions are given by van Dishoeck \& Blake
(1998), Hartquist et al.\ (1998), van Dishoeck \& Hogerheijde (1999)
and Langer et al.\ (2000).  Schilke et al.\ (this volume) present high
spatial resolution interferometer studies, whereas Macdonald \&
Thompson (this volume) focus on submillimeter data of hot
core/ultracompact H~II regions. Ices are discussed by Ehrenfreund \&
Schutte (this volume).

\begin{figure}[t]
\plotfiddle{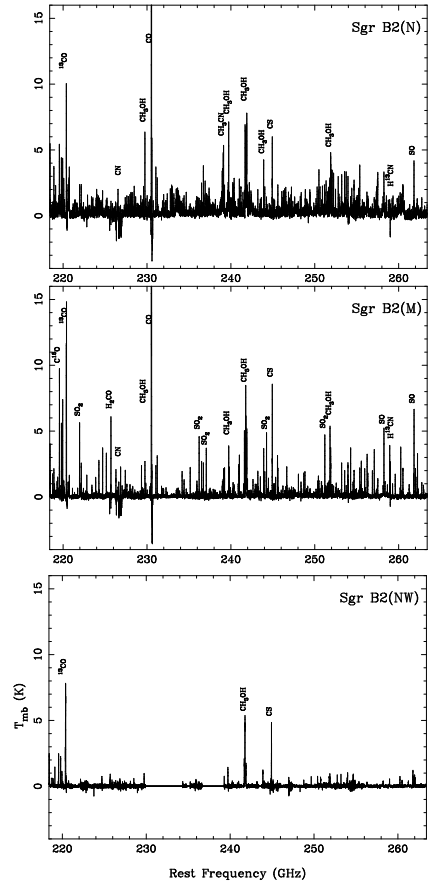}{11.0cm}{0}{50}{50}{-150}{-30}
\caption{Overview of the SEST 230 GHz line survey of SgrB2 at the N, M and
NW positions, illustrating the chemical differentiation in high-mass
star-forming regions (from:
Nummelin et al.\ 1998).}
\label{fig-1}
\vspace{-0.3cm}
\end{figure}

\section{Submillimeter and Infrared Diagnostics}

The majority of molecules are detected at (sub-)millimeter
wavelengths, and line surveys highlight the large variations in
chemical composition between different YSOs, both within the same
parent molecular cloud and between different clouds.  The recent 1--3
mm surveys of Sgr~B2 (Nummelin et al.\ 1998, Ohishi \& Kaifu 1999)
dramatically illustrate the strong variations between various
positions (see Figure~1). The North position is typical of
`hot core'-type spectra, which are rich in lines of saturated
organic molecules. This position has also been named 
the `large molecule heimat'
(e.g., Kuan \& Snyder 1994, Liu \& Snyder 1999).  The Middle position
has strong SO and SO$_2$ lines, whereas the Northwest position has a
less-crowded spectrum with lines of ions and long carbon chains.  A
similar differentiation has been observed for three positions in the
W~3 giant molecular cloud by Helmich \& van Dishoeck (1997), who
suggested an evolutionary sequence based on the chemistry.

\begin{figure}[t]
\plotfiddle{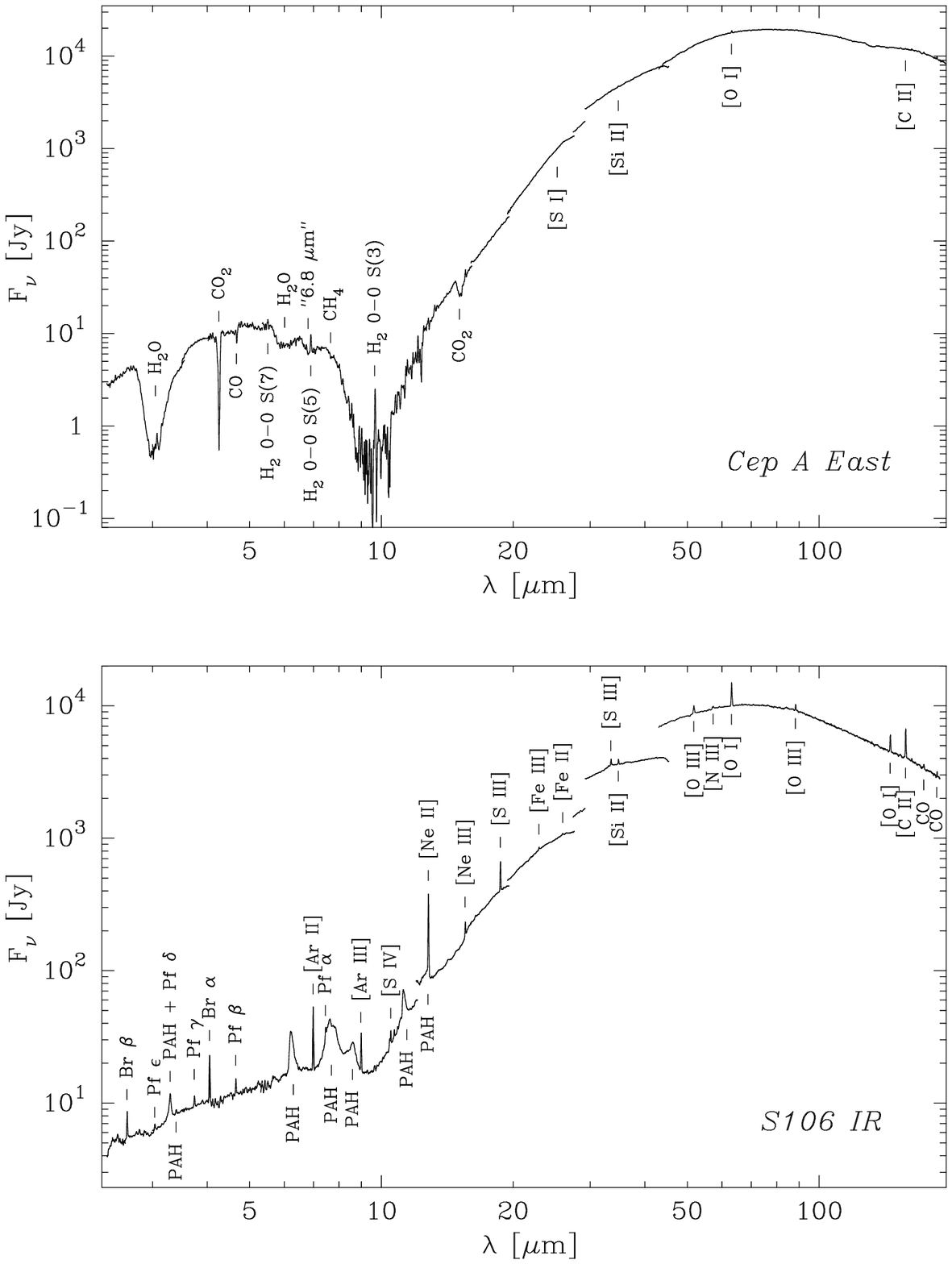}{10cm}{0}{55}{55}{-150}{-115}
\caption{ISO-SWS spectra of two massive YSOs at different evolutionary stages.
Cep A ($L \approx 2.4\times 10^4$ L$_{\odot}$) is in the embedded
stage, whereas S~106 ($L \approx 4.2\times 10^4$ L$_{\odot}$) is
in a more evolved stage (from:
van den Ancker et al.\ 2000a).}
\label{fig-2}
\end{figure}

The availability of complete infrared spectra from 2.4--200 $\mu$m
with the {\it Infrared Space Observatory }(ISO) allows complementary
variations in infrared features to be studied.  Figure~2 shows an
example of ISO--SWS and LWS spectra of two objects: Cep A ($L \approx
2.4\times 10^4$ L$_{\odot}$) and S~106 ($L \approx 4.2\times 10^4$
L$_{\odot}$). The Cep A spectrum is characteristic of the deeply
embedded phase, in which the silicates and ices in the cold envelope
are seen in absorption.  The S~106 spectrum is typical of a more
evolved massive YSO, with strong atomic and ionic lines in emission
and prominent PAH features.  A similar sequence has been shown by
Ehrenfreund et al.\ (1998) for a set of southern massive young stars
with luminosities up to $4\times 10^5$ L$_{\odot}$.

The most successful models for explaining these different chemical
characteristics involve accretion of species in an icy mantle during the
(pre-)collapse phase, followed by grain-surface chemistry and evaporation of
ices once the YSO has started to heat its surroundings (e.g., Millar 1997). 
The evaporated molecules subsequently drive a rapid high-temperature gas-phase
chemistry for a period of $\sim 10^4 - 10^5$ yr, resulting in complex,
saturated organic molecules (e.g., Charnley et al.\ 1992, 1995; Charnley 1997;
Caselli et al.\ 1993, Viti \& Williams 1999). The abundance ratios of species
such as CH$_3$OCH$_3$/CH$_3$OH and SO$_2$/H$_2$S show strong variations with
time, and may be used as `chemical clocks' for a period of 5000--30,000 yr
since evaporation. Once most of the envelope has cleared, the ultraviolet
radiation can escape and forms a photon-dominated region (PDR) at the
surrounding cloud material, in which molecules are dissociated into radicals
(e.g., HCN $\to$ CN) and PAH molecules excited to produce infrared emission.
The (ultra-)compact H~II region gives rise to strong ionic lines due to
photoionization.

\begin{table}[t]
\caption{Chemical characteristics of massive star-forming regions}
\begin{center}\scriptsize
\begin{tabular}{lllll}
\hline
\noalign{\smallskip}
Component & Chemical & Submillimeter & Infrared & Examples \\
          & characteristics & diagnostics & diagnostics \\
\tableline
\noalign{\smallskip}
Dense cloud & Low-T chemistry &Ions, long-chains &Simple ices & SgrB2 (NW) \\
            &                 & (HC$_5$N, ...)   & (H$_2$O, CO$_2$) \\
\noalign{\smallskip}
Cold envelope & Low-T chemistry, & Simple species & Ices & N7538 IRS9, \\
              & Heavy depletions  & (CS, H$_2$CO) & (H$_2$O, CO$_2$, CH$_3$OH)
     & W~33A \\
\noalign{\smallskip}
Inner warm  & Evaporation  & High T$_{\rm ex}$ & High gas/solid, High
      & GL 2591, \\
envelope    &      & (CH$_3$OH) & T$_{\rm ex}$, Heated ices  
                  & GL 2136 \\
       & & & (C$_2$H$_2$, H$_2$O, CO$_2$) \\
\noalign{\smallskip}
Hot core & High-T chemistry  & Complex organics & Hydrides & Orion hot core, \\
      &         & (CH$_3$OCH$_3$, CH$_3$CN, & (OH, H$_2$O) 
& SgrB2(N),G34.3 \\
&&vib.\ excited mol.) && W~3(H$_2$O) \\
\noalign{\smallskip}
Outflow: & Shock chemistry, & Si- and S-species & 
   Atomic lines, Hydrides & W~3 IRS5,  \\
Direct impact      &Sputtering & (SiO, SO$_2$) &([S I], H$_2$O) & SgrB2(M) \\
\noalign{\smallskip}
PDR, Compact & Photodissociation,& Ions, radicals & Ionic lines, PAHs
& S~140, \\
H~II regions & Photoionization & (CN/HCN, CO$^+$) &([NeII], [CII])
& W~3 IRS4 \\
 
\tableline
\end{tabular}
\end{center}
\end{table}

Table~1 summarizes the chemical characteristics of the
various physical components, together with the observational
diagnostics at submillimeter and infrared wavelengths. 
Within the single-dish submillimeter and ISO beams,
many of these components are blended together and interferometer
observations will be essential to disentangle them.  Nevertheless, the
single-dish data are useful because they encompass the entire envelope
and highlight the dominant component in the beam. Combined
with the above chemical scenario, one may then attempt to establish an
evolutionary sequence of the sources.

The physical distinction between the `hot core' and the warm
inner envelope listed in Table~1 is currently not clear: does the `hot
core' represent a separate physical component or is it simply the
inner warm envelope at a different stage of chemical evolution? Even
from an observational point of view, there appear to be different
types of `hot cores': some of them are internally heated by the young
star (e.g., W~3(H$_2$O)), whereas others may just be dense clumps of
gas heated externally (e.g., the Orion compact ridge).  This point
will be further discussed in \S\S 4 and 5.  Disks are not included in
Table~1, because little is known about their chemical characteristics,
or even their existence, around high-mass YSOs (see Norris, this
volume).

\section{Intermediate-Mass YSOs}

Intermediate-mass pre-main sequence stars, in particular the so-called
Herbig Ae/Be stars, have received increased observational attention in
recent years (see Waters \& Waelkens 1998 for a review).  These stars
have spectral type A or B and show infrared excesses due to 
circumstellar dust. Typical luminosities are in the range $10^2 -
10^4$ L$_{\odot}$, and several objects have been located within 1 kpc
distance. Systematic mapping of CO and the submillimeter continuum of
a sample of objects has been performed by Fuente et al.\ (1998) and
Henning et al.\ (1998). The data show the dispersion of the envelope
with time starting from the deeply embedded phase (e.g.,
LkH$\alpha$234) to the intermediate stage of a PDR (e.g., HD 200775
illuminating the reflection nebula NGC 7023) to the more evolved stage
where the molecular gas has disappeared completely (e.g., HD 52721).
The increasing importance of photodissociation in the chemistry is
probed by the increase in the CN/HCN abundance ratio.  This ratio has
been shown in other high-mass sources to be an excellent tracer of
PDRs (e.g., Simon et al. 1997, Jansen et al.\ 1995).  Line surveys of
these objects in selected frequency ranges would be useful to
investigate their chemical complexity, especially in the embedded
phase.

ISO-SWS observations of a large sample of Herbig Ae/Be stars have been
performed by van den Ancker et al.\ (2000b,c). In the embedded phase,
shock indicators such as [S I] 25.2 $\mu$m are strong, whereas in the
later phases PDR indicators such as PAHs are prominent.  An excellent
example of this evolutionary sequence is provided by three Herbig Ae
stars in the BD+40$^o$4124 region ($d\approx 1$ kpc).  The data
suggest that in the early phases, the heating of the envelope is
dominated by shocks, whereas in later phases it is controlled by
ultraviolet photons.

\begin{figure}[t]
\vspace{-3.8cm}
\plottwo{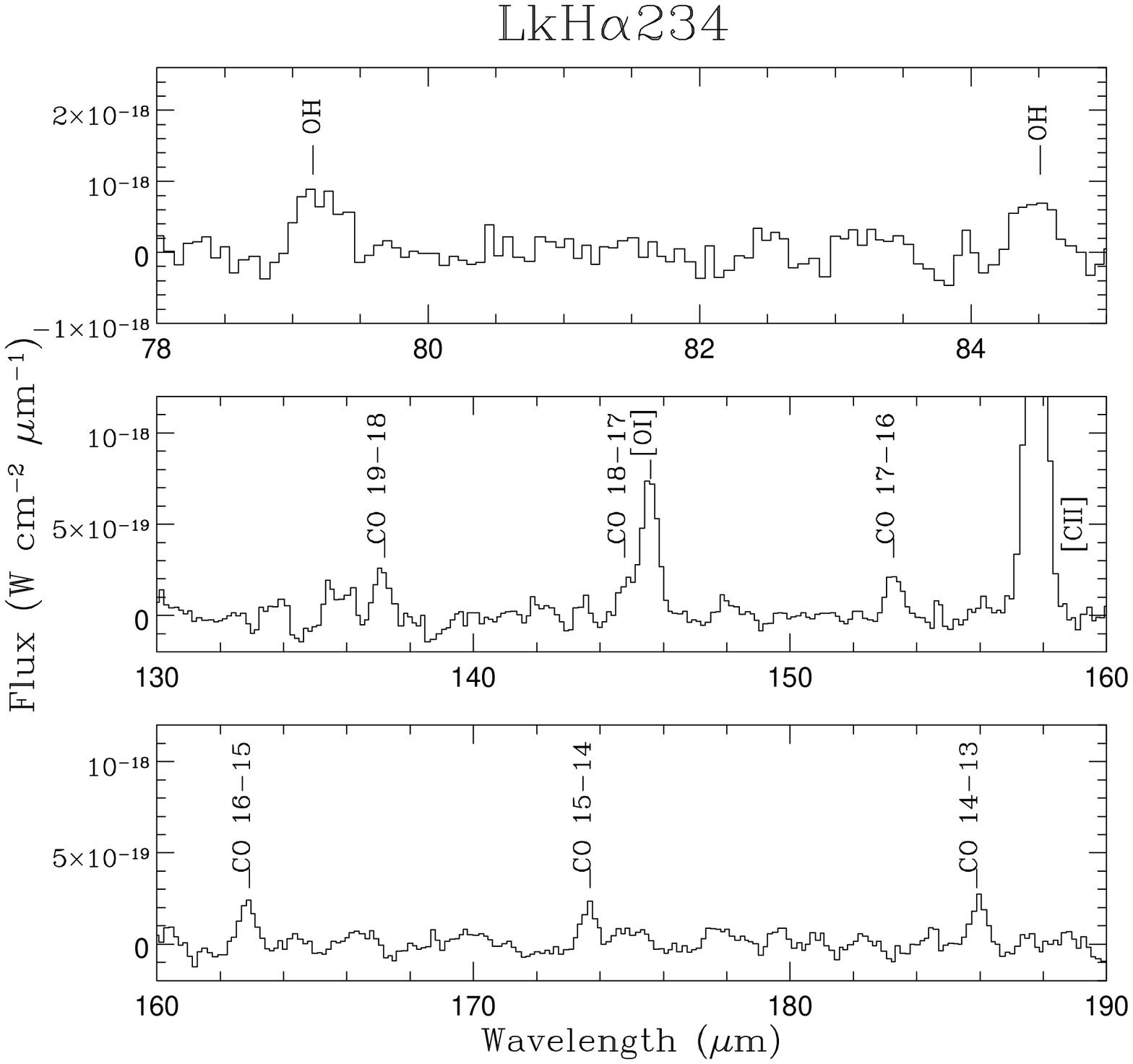}{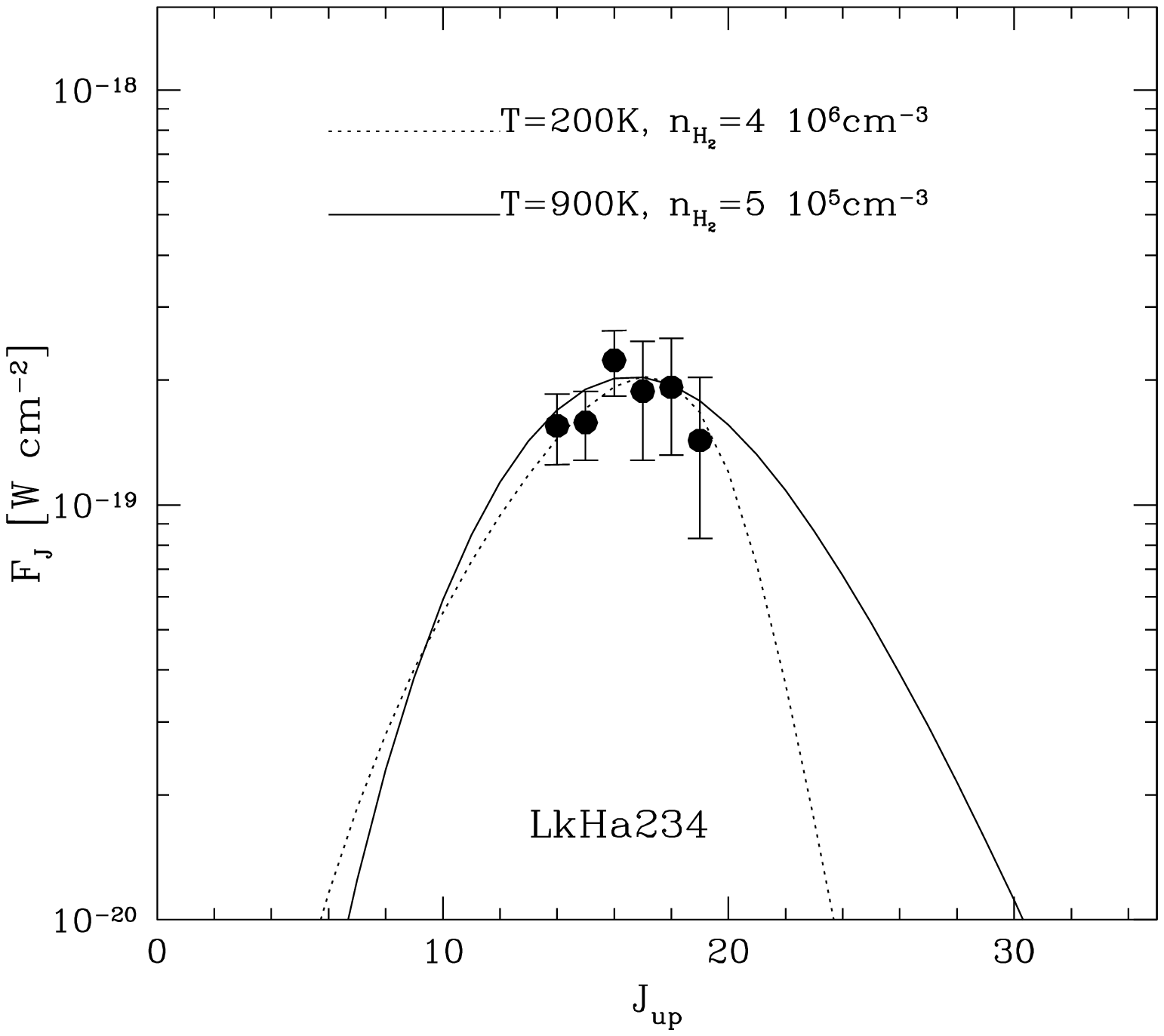}
\vspace{-0.3cm}
\caption{Left: ISO-LWS spectra of the intermediate mass YSO 
LkH$\alpha$234, showing strong atomic fine-structure lines of
[C II] 158 $\mu$m and [O~I] 63 $\mu$m, but weak molecular CO and
OH lines. Right: CO excitation diagram, indicating the presence
of a warm ($T\approx 200-900$~K), dense ($n>10^5$ cm$^{-3}$) 
region (from:
Giannini et al.\ 2000).}
\label{fig-3}
\vspace{-0.5cm}
\end{figure}

ISO-LWS data have been obtained for a similar sample of Herbig Ae/Be
stars by Lorenzetti et al.\ (1999) and Giannini et al.\ (2000), and
are summarized by Saraceno et al.\ (1999). The [C II] 158 $\mu$m and
[O I] 63 and 145 $\mu$m lines are prominent in many objects and are
due primarily to the PDR component in the large LWS beam ($\sim
80''$).  High-$J$ CO and OH far-infrared lines have been detected in
some objects and indicate the presence of a compact, high temperature
and density region of $\sim$1000 AU in size, presumably tracing the
inner warm envelope (see Figure~3).
Far-infrared lines of H$_2$O are seen in low-mass YSO spectra, but are
weak or absent in those of intermediate- and high-mass YSOs, with the
exception of Orion-KL and SgrB2 (e.g., Harwit et al.\ 1998, Cernicharo
et al.\ 1997, Wright et al.\ 2000).  The absence of H$_2$O lines in
higher-mass objects may be partly due to the larger distance of these
objects, resulting in substantial dilution in the LWS beam. However,
photodissociation of H$_2$O to OH and O by the enhanced ultraviolet radiation
may also play a role.

In summary, both the submillimeter and infrared diagnostics reveal an
evolutionary sequence from the youngest `Group I' objects to `Group
III' objects (cf.\ classification by Fuente et al.\ 1998), in which
the envelope is gradually dispersed. Such a sequence is analogous to
the transition from embedded Class 0/I objects to more evolved Class
II/III objects in the case of low-mass stars (Adams et al.\ 1987).  The
ISO data provide insight into the relative importance of the heating
and removal mechanisms of the envelope.  At the early stages of
intermediate-mass star formation, shocks due to outflows appear to
dominate whereas at later stages radiation is more important.

\section{Embedded, Infrared-Bright Massive YSOs}

\subsection{Sample}

The availability of complete, high quality ISO spectra for a
significant sample of massive young stars provides a unique
opportunity to study these sources through a combination of infrared
and submillimeter spectroscopy, and further develop these diagnostics.
Van der Tak et al.\ (2000a) have selected a set of $\sim$10 deeply
embedded massive YSOs which are bright at mid-infrared wavelengths (12
$\mu$m flux $>$ 100 Jy), have luminosities of $10^3 - 2\times 10^5$
L$_{\odot}$ and distances $d\leq$4 kpc. The sources are all in an
early evolutionary state (comparable to the `Class 0/I' or `Group I'
stages of low- and intermediate-mass stars), as indicated by their
weak radio continuum emission and absence of ionic lines and PAH
features.  In addition to ISO spectra, JCMT submillimeter data and
OVRO interferometer observations have been obtained.  For most of the
objects high spectral resolution ground-based infrared data of CO,
$^{13}$CO and H$_3^+$ are available (Mitchell et al.\ 1990, Geballe \&
Oka 1996, McCall et al.\ 1999), and occasionally H$_2$ (Lacy et
al.\ 1994, Kulesa et al.\ 1999). For comparison, 5 infrared-weak
sources with similar luminosities are studied at submillimeter
wavelengths only. This latter set includes hot cores and ultracompact
H~II regions such as W~3(H$_2$O), IRAS 20126+4104 (Cesaroni et al.\ 
1997, 1999), and NGC~6334 IRS1.

\subsection{Physical structure of the envelope}

In order to derive molecular abundances from the observations, a good
physical model of the envelope is a prerequisite. Van der Tak et al.\ 
(1999, 2000a) outline the techniques used to constrain the temperature
and density structure (Figure~4). The total mass within the
beam is derived from submillimeter photometry, whereas the size scale
of the envelope is constrained from line and continuum maps.  The dust
opacity has been taken from Ossenkopf \& Henning (1994) and yields
values for the mass which are consistent with those derived from
C$^{17}$O for warm sources where CO is not depleted
onto grains.

\begin{figure}[t]
\plotfiddle{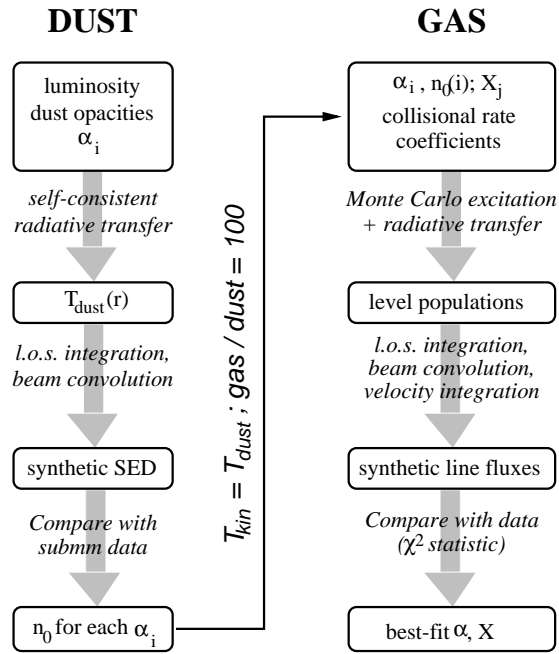}{7.7cm}{0}{45}{45}{-120}{-60}
\caption{Overview of method used for constraining the temperature and density
structure of envelopes. A power-law density structure with trial values
for the exponent
$\alpha_i$ and molecular abundances $X_j$ is adopted
(based on: van der Tak et al.~2000a).}
\label{fig-4}
\end{figure}

The temperature structure of the dust is calculated taking the
observed luminosity of the source, given a power-law density structure
(see below). At large distances from the star, the temperature follows
the optically thin relation $\propto r^{-0.4}$, whereas at smaller
distances the dust becomes optically thick at infrared wavelengths and
the temperature increases more steeply (see Figure~5). It is assumed
that $T_{\rm gas}=T_{\rm dust}$,  consistent with explicit calculations
of the gas and dust temperatures by, e.g., Doty \& Neufeld (1997)
for these high densities.

The continuum data are sensitive to temperature and column density,
but not to density.  Observations of a molecule with a large dipole
moment are needed to subsequently constrain the density structure. One
of the best choices is CS and its isotope C$^{34}$S, for which lines
ranging from $J$=2--1 to 10--9 have been observed. Assuming a
power-law density profile $n(r)= n_o (r/r_o)^{-\alpha}$, values of
$\alpha$ can be determined from minimizing $\chi^2$ between the CS
line data and excitation models. The radiative transfer in the lines
is treated through a Monte-Carlo method.  The best fit to the data on
the infrared-bright sources is obtained for $\alpha = 1.0 - 1.5$,
whereas the hot core/compact H~II region sample requires higher
values, $\alpha \approx 2$. This derivation assumes that the CS
abundance is constant through the envelope; if it increases with
higher temperatures, such as may be the case for hot cores, the values
of $\alpha$ are lowered.  Note that the derived values of
$\alpha=1.0-1.5$ are lower than those found for deeply embedded
low-mass objects, where $\alpha\approx 2$ (e.g., Motte et al.\ 1998,
Hogerheijde et al.\ 1999).

\begin{figure}[t]
\plotfiddle{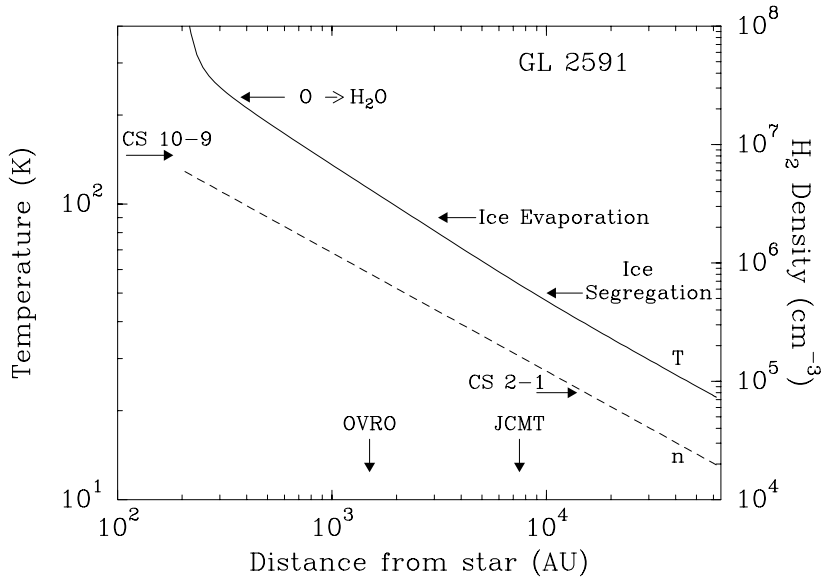}{7.0cm}{-90}{45}{45}{-170}{260}
\vspace{-0.3cm}
\caption{Derived temperature and density structure for the envelope
around the massive YSO
GL 2591 (L=$2\times 10^4$ L$_{\odot}$, $d$=1 kpc). 
The critical densities of the CS 2--1 and 10--9 lines are
indicated, as are the typical beam sizes of the JCMT 
at 345 GHz and OVRO at 100 GHz (based on: van der Tak et 
al.\ 1999).}
\label{fig-5}
\vspace{-0.5cm}
\end{figure}

Figure 5 displays the derived temperature and density structure
for the source GL 2591, together with the sizes of the JCMT and
OVRO beams. While the submillimeter data are weighted toward the
colder, outer envelope, the infrared absorption line observations sample
a pencil-beam line of sight toward the YSO and are more sensitive to
the inner warm ($\sim 1000$~K) region. On these small scales, the
envelope structure deviates from a radial power law, which decreases
the optical depth at near-infrared wavelengths by a factor of 
$\sim 3$ (van der Tak et al.\
1999).

For sources for which interferometer data are available, unresolved
compact continuum emission is detected on scales of a few thousand AU
or less.  This emission is clearly enhanced compared with that
expected from the inner ``tip'' of the power-law envelope, and its
spectral index indicates optically thick warm dust, most likely in a
dense circumstellar shell or disk.  The presence of this shell or disk
is also indicated by the prevalence of blue-shifted outflowing dense
gas without a red-shifted counterpart on $< 10''$ scales.

\subsection{Chemical structure: infrared absorption lines}

The ISO-SWS spectra of the infrared-bright sources show absorption by
various gas-phase molecules, in addition to strong features by ices.
Molecules such as CO$_2$ (van Dishoeck et al.\ 1996,
Boonman et al.\ 1999, 2000a), H$_2$O (van Dishoeck \& Helmich 1996, Boonman
et al.\ 2000b), CH$_4$ (Boogert et al.\ 1998), HCN and C$_2$H$_2$
(Lahuis \& van Dishoeck 2000) have been detected (see also van
Dishoeck 1998, Dartois et al.\ 1998).

In the infrared, absorption out of all $J$-levels is observed in a
single spectrum. The excitation temperatures $T_{\rm ex}$ of the
various molecules, calculated assuming LTE, range from $\ltsim 100$ to
$\sim 1000$~K between sources, giving direct information on 
the physical component in which the molecules reside. While CO 
is well-mixed throughout the envelopes, H$_2$O, HCN and C$_2$H$_2$ are
enhanced at high temperatures. In contrast, CO$_2$ seems to avoid the
hottest gas. High spectral resolution ground-based data of HCN and
C$_2$H$_2$ by Carr et al.\ (1995) and Lacy et al.\ (1989) for a few
objects suggest line widths of at most a few km s$^{-1}$, excluding an
origin in outflowing gas.

The abundances of H$_2$O, HCN and C$_2$H$_2$ increase by factors of
$\gtsim 10$ with increasing $T_{\rm ex}$ (see Figure~7). The warm
H$_2$O must be limited to a $\ltsim 1000$~AU region, since the pure
rotational lines are generally not detected in the $80''$ ISO-LWS beam
(Wright et al.\ 1997).  For CO$_2$, the abundance variation between
sources is less than a factor of 10, and no clear trend with $T_{\rm
  ex}$ is found. For the same sources, the H$_2$O and CO$_2$ ice
abundances show a decrease by an order of magnitude, consistent with
evaporation of the ices. However, the gas-phase H$_2$O and CO$_2$
abundances are factors of $\sim 10$ lower than expected if all
evaporated molecules stayed in the gas phase, indicating that
significant chemical processing occurs after evaporation. More
detailed modeling using the source structures derived from
submillimeter data is in progress.

\begin{figure}[t]
\plotfiddle{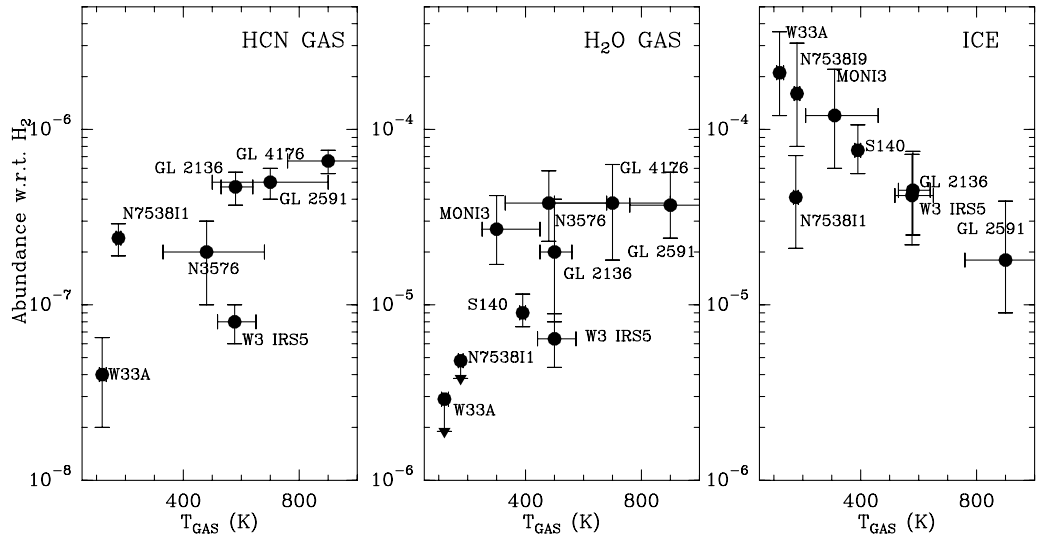}{6.8cm}{-90}{50}{50}{-210}{260}
\caption{{\it Left:}~Abundance of HCN,
$N$(HCN)/$N_{\rm tot}$(H$_2$), as a function of the
$^{13}$CO excitation temperature measured by
Mitchell et al.\ (1990). {\it Middle:} Abundance of H$_2$O in the warm gas
$N$(H$_2$O)/$N_{\rm warm}$(H$_2$) as a function of $^{13}$CO excitation
temperature;
{\it Right:}~Abundance of H$_2$O ice as a function of
$^{13}$CO excitation temperature. The same trend
is found if the dust temperature, as derived from the 45/100 $\mu$m
flux, is used (based on: Lahuis \& van Dishoeck 2000; van Dishoeck 1998).}
\label{fig-6}
\end{figure}

\subsection{Chemical structure: submillimeter emission}

The JCMT data of the infrared-bright objects show strong lines, but
lack the typical crowded `hot core' spectra observed for objects such
as W~3(H$_2$O) and NGC 6334 IRS1. Complex organics such as
CH$_3$OCH$_3$ and CH$_3$OCHO are detected in some sources (e.g., GL
2591, NGC 7538 IRS1), but are not as prominent as in the comparison
sources. Yet warm gas is clearly present in these objects.
Is the `hot core' still too small to be picked up by the
single dish beams, or are the abundances of these molecules not (yet)
enhanced?

To investigate this question, 
van der Tak et al.\ (2000b) consider the analysis of
two species, H$_2$CO and CH$_3$OH. Both
species have many lines throughout the submillimeter originating from
low- and high-lying energy levels.  Given the physical structure
determined in \S 4.2, abundance {\it profiles} can be constrained. Two
extreme, but chemically plausible models are considered: (i) a model
with a constant abundance throughout the envelope. This model is
motivated by the fact that pure gas-phase reaction schemes do not show
large variations in calculated abundances between 20 and 100~K; (ii) a
model in which the abundance `jumps' to a higher value at the ice
evaporation temperature, $T_d\approx 90$~K. In this model, the
abundances in the outer envelope are set at those observed in cold
clouds, so that the only free parameter is the amount of abundance
increase.

It is found that the H$_2$CO data can be well fit with a constant
abundance of a few $\times 10^{-9}$ throughout the envelope. However,
the high $J,K$ data for CH$_3$OH require a jump in its abundance from
$\sim 10^{-9}$ to $\sim 10^{-7}$ for the warmer sources.  This is
consistent with the derived excitation temperatures: H$_2$CO has a
rather narrow range of $T_{\rm ex}$=50--90~K, whereas CH$_3$OH shows
$T_{\rm ex}$=30--200~K. Moreover, the interferometer maps of CH$_3$OH
rule out constant abundance models. The jump observed for CH$_3$OH is
chemically plausible since this molecule is known to be present in icy
grain mantles with abundances of 5--40\% with respect to H$_2$O ice,
i.e., $\sim 10^{-7}-10^{-6}$ w.r.t.\ H$_2$.  Similar increases in the
abundances of organic molecules (e.g., CH$_3$OH, C$_2$H$_3$CN, ....)
are found with increasing $T_{\rm dust}$ for a set of `hot--core'
objects by Ikeda \& Ohishi (1999).

\subsection{Comparison with chemical models}

Both the infrared and submillimeter data show increases in the
abundances of various molecules with increasing temperature. Four
types of species can be distinguished: (i) `passive' molecules which
are formed in the gas phase, freeze out onto grains during the cold
(pre-)collapse phase and are released during warm-up without chemical
modification (e.g., CO, C$_2$H$_2$); (ii) molecules which are formed
on the grains during the cold phase by surface reactions and are
subsequently released into the warm gas (e.g., CH$_3$OH); (iii)
molecules which are formed in the warm gas by gas-phase reactions with
evaporated molecules (e.g., CH$_3$OCH$_3$); (iv) molecules which are
formed in the hot gas by high temperature reactions (e.g., HCN). These
types of molecules
are associated with characteristic temperatures of (a) $T_{\rm
  dust}<20$~K, where CO is frozen out; the presence of CO ice is
thought to be essential for the formation of CH$_3$OH; (b) $T_{\rm
  dust} \approx 90$~K, where all ices evaporate on a time scale of
$<10^5$ yr; and (c) $T_{\rm gas}>230$~K, where gas-phase reactions
drive the available atomic oxygen into water through the reactions O +
H$_2$ $\to$ OH $\to$ H$_2$O (Ceccarelli et al.\ 1996, Charnley 1997).
Atomic oxygen is one of the main destroyers of radicals and carbon
chains, so that its absence leads to enhanced abundances of species
like HCN and HC$_3$N in hot gas.

Water is abundantly formed on the grains, but the fact that the
H$_2$O abundance in the hot gas is not as large as that of the ices
suggests that H$_2$O is broken down to O and OH after evaporation by
reactions with H$_3^+$. H$_2$O can subsequently be reformed in warm
gas at temperatures above $\sim$230~K, but Figure~7 indicates that not
all available gas-phase oxygen is driven into H$_2$O, as the models
suggest.  The low abundance of CO$_2$ in the warm gas is still a
puzzle, since evaporation of abundant CO$_2$ ice is observed. The
molecule must be broken down rapidly in the warm gas, with no
reformation through the CO + OH $\to$ CO$_2$ reaction (see question by
Minh).

\subsection{Evolution?}

The objects studied by van der Tak et al.\ (2000a) are all in an early
stage of evolution, when the young stars are still deeply embedded in
their collapsing envelope.  Nevertheless, even within this narrow
evolutionary range, there is ample evidence for physical and chemical
differentiation of the sources.  This is clearly traced by the
increase in the gas/solid ratios, the increase in abundances of several
molecules, the decrease in the ice abundances, and the increase
of the amount of crystalline ice with increasing temperature
(Boogert et al.\ 2000a).

The fact that the various indicators involve different
characteristic temperatures ranging from $<$50 K (evaporation of
apolar ices) to 1000 K ($T_{\rm ex}$ of gas-phase molecules)
indicates that the heating is not a local effect, but that `global
warming' occurs throughout the envelope. Moreover, it cannot be a
geometrical line-of-sight effect in the mid-infrared data, since the
far--infrared continuum (45/100 $\mu$m) and submillimeter line data
(CH$_3$OH) show the same trend. Shocks with different filling factors
are excluded for the same reason.

Can we relate this `global warming' of the envelope to an evolutionary
effect, or is it determined by other factors?  The absence of a
correlation of the above indicators with luminosity or mass of the
source argues against them being the sole controlling factor.  The
only significant trend is found with the ratio of envelope mass over
stellar mass.  The physical interpretation of such a relation would be
that with time, the envelope is dispersed by the star, resulting in a
higher temperature throughout the envelope.

\section{Outstanding questions and future directions}

The results discussed here suggest that the observed chemical abundance and
temperature variations can indeed be used to trace the evolution of
the sources, and that, as in the case of low- and intermediate-mass
stars, the dispersion of the envelope plays a crucial role. 
The combination of infrared and submillimeter diagnostics
is very important  in the analysis. An
important next step would be to use these diagnostics to probe a much
wider range of evolutionary stages for high-mass stars, especially in
the hot core and (ultra-)compact H II region stages, to develop a more
complete scenario of high-mass star formation.  The relation between
the inner warm envelope and the `hot core' is still uncertain: several
objects have been observed which clearly have hot gas and evaporated
ices (including CH$_3$OH) in their inner regions, but which do not
show the typical crowded `hot core' submillimeter spectra. Are these
objects just on their way to the `hot core' chemical phase?  Or is the
`hot core' a separate physical component, e.g., a dense shell at the
edge of the expanding hyper-compact H~II region due to the pressure
from the ionized gas, which is still too small to be
picked up by the single-dish beams?  In either case, time or evolution
plays a role and would constrain the ages of the infrared-bright
sources to less than a few $\times 10^4$ yr since evaporation.
Interferometer data provide evidence for the presence of a separate
physical component in the inner 1000 AU, but lack the spatial
resolution to distinguish a shell from any remnant disk, for example
on kinematic grounds.  

An important difference between high- and low-mass objects may be the
mechanism for the heating and dispersion of their envelopes.  For
low-mass YSOs, entrainment of material in outflows is the
dominant process (Lada 1999).  For intermediate-mass stars, outflows
are important in the early phase, but ultraviolet radiation
becomes dominant in the later stages (see \S~3).  The situation for
high-mass stars is still unclear.  The systematic increase in
gas/solid ratios and gas-phase abundances point to global heating of
the gas and dust, consistent with a radiative mechanism. However, a
clear chemical signature of ultraviolet radiation on gas-phase species
and ices in the embedded phase has not yet been identified, making it
difficult to calibrate its effect.  On the other hand, high-mass stars
are known to have powerful outflows and winds, but a quantitative
comparison between their effectiveness in heating an extended part
of the envelope and removing material is still
lacking. Geometrical effects are more important in less embedded
systems, as is the case for low-mass stars, where the circumstellar
disk may shield part of the envelope from heating (Boogert et al.\ 
2000b).

To what extent does the chemical evolution picture also apply to
low-mass stars?  Many of the chemical processes and characteristics
listed in Table~1 are also known to occur for low-mass YSOs, but
several important diagnostic tools are still lacking.  In particular,
sensitive mid-infrared spectroscopy is urgently needed to trace the evolution
of the ices for low-mass YSOs and determine gas/solid ratios.  Also,
molecules as complex as CH$_3$OCH$_3$ and C$_2$H$_5$CN have not yet
been detected toward low-mass YSOs, although the limits are not very
stringent (e.g., van Dishoeck et al.\ 1995). Evaporation of ices
clearly occurs in low-mass environments as evidenced by enhanced
abundances of grain-surface molecules in shocks (e.g., Bachiller \&
P\'erez-Guti\'errez 1997), but whether a similar `hot core' chemistry
ensues is not yet known.  Differences in the 
H/H$_2$ ratio and temperature structure in the (pre-)collapse
phase may affect the
grain-surface chemistry and the ice composition,
leading to different abundances of solid CH$_3$OH, which 
is an essential ingredient
for building complex molecules.

Future instrumentation with high spatial resolution ($< 1''$) and
high sensitivity will be essential to make progress in our
understanding of the earliest phase of massive star formation, in
particular the SMA and ALMA at submillimeter wavelengths, and SIRTF,
SOFIA, FIRST and ultimately NGST at mid- and far-infrared wavelengths.

\acknowledgments

The authors are grateful to G.A. Blake, A.C.A.\ Boogert, A.M.S.\ Boonman, P.\ 
Ehrenfreund, N.J.\ Evans, T.\ Giannini, F.\ Lahuis, L.G.\ Mundy, A.\ 
Nummelin, W.A.\ Schutte, A.G.G.M.\ Tielens, and M.E.\ van den Ancker
for discussions, collaborations and figures. This work was supported
by NWO grant 614.41.003

\begin{question}{M.\ Ohishi}
I agree with the point you mentioned, that the chemical differences among
hot cores is due to a difference of evolutionary stage. Now we have
several well-known hot cores such as Orion KL/S, W~3 IRS5/H$_2$O/IRS4,
SgrB2 N/M/NW etc. Can you give us your personal view on the evolutionary
differences of these sources?
\end{question} 

\begin{answer}{E.F.\ van Dishoeck}
Van der Tak et al.\ (2000a) argue that the infrared-bright objects such as W~3
IRS5 represent an earlier evolutionary phase than the hot cores, on the basis
of an anti-correlation with the radio continuum. The physical picture is that
the ionizing UV radiation and stellar winds push the hottest dust in the inner
regions further out, decreasing the temperature of the dust and thus the
near-infrared continuum. At the same time, the size of the region which can be
ionized is increased. The `erosion' of the envelopes thus occurs from the
inside out. For other sources, infrared diagnostics are lacking, so that the
situation is less clear. It would be great if chemistry could help to tie down
the time scales of the various phases.
\end{answer}

\begin{question}{W.\ Irvine}
How do you interpret the behavior of the PAH features as a function of
evolutionary stage in the sources that you discussed?
\end{question}

\begin{answer}{E.F.\ van Dishoeck}
The absence of PAH features in the early embedded stage can be due either
to a lack of ultraviolet radiation to excite the features or to an absence
of the carriers. Manske \& Henning (1999, A\&A 349, 907) have argued 
for the case of Herbig Ae/Be stars that
there should be sufficient radiation to excite PAHs in the envelope/disk
system,
so that the lack is likely due to the absence of the PAHs
themselves. Perhaps the PAHs have accreted into the icy mantles at the
high densities in the inner envelope and do not evaporate
and/or are chemically transformed into
other more refractory species on grains. Alternatively, the region producing
ultraviolet radiation (H~II region) may be very small 
in these massive objects, and the photons may not reach the PAH-rich
material or have a very small beam filling factor.
Once the envelope breaks up and
ultraviolet radiation can escape to the less dense
outer envelope, the PAH features from those
regions will appear in spectra taken with large beams.
\end{answer}

\begin{question}{T.\ Geballe}
You said that there is little evidence of the effect of ultraviolet radiation
on solid-state chemistry. Isn't the 4.6 $\mu$m XCN feature a good example
of that influence?
\end{question}

\begin{answer}{E.F.\ van Dishoeck}
The `XCN' feature is indeed the best candidate for tracing the 
ultraviolet processing
of ices. If ascribed to OCN$^-$,  it likely involves HNCO as
a precursor. In the laboratory, HNCO is produced by photochemical
reactions of CO and NH$_3$, but in the interstellar medium grain surface
chemistry is an alternative possibility which does not necessarily
involve ultraviolet radiation (see Ehrenfreund \& Schutte,
this volume). 
\end{answer}

\begin{question}{Y.C.\ Minh}
Do you have an explanation of the low and constant abundances of CO$_2$
in the gas phase?
\end{question}

\begin{answer}{E.F.\ van Dishoeck}
Charnley \& Kaufman (2000, ApJ, 529, L111) argue that the evaporated CO$_2$ is
destroyed by reactions with atomic hydrogen at high temperatures in shocks.
This is an interesting suggestion, but needs to be tested against other
species such as H$_2$O and H$_2$S which can be destroyed by
reactions with atomic hydrogen as well. Also, the amount of material 
in the envelope that can be affected by shocks is not clear.
\end{answer}

\end{document}